\documentclass[12pt]{article}
\usepackage{setspace}
\usepackage{amssymb}
\usepackage{amsmath}
\usepackage{wrapfig}
\usepackage{graphicx}
\usepackage{hyperref}
\usepackage{authblk}

\begin{document}
\title{ Statistical Theory of Breakup Reactions}
%
%

\author{Carlos A. Bertulani$^{1}$\thanks{carlos.bertulani@tamuc.edu},  Pierre Descouvemont$^{2}$\thanks{pdesc@ulb.ac.be}, Mahir S. Hussein$^{3}$\thanks{hussein@if.usp.br}
}

\affil{\small Department of Physics and Astronomy, Texas A\&M University-Commerce, Commerce,
TX 75429, USA 
\\
          Physique Nucl\'eaire Th\'eorique et Physique Math\'ematique, C.P. 229,
Universit\'e Libre de Bruxelles (ULB), B 1050 Brussels, Belgium 
\\
           Instituto de Estudos Avan\c{c}ados, Universidade de S\~{a}o Paulo C. P.
72012, 05508-970 S\~{a}o Paulo-SP, Brazil, and Instituto de F\'{\i}sica,
Universidade de S\~{a}o Paulo, C. P. 66318, 05314-970 S\~{a}o Paulo,-SP, Brazil
          }
\maketitle

\abstract{%
We propose alternatives to coupled-channels calculations with loosely-bound exotic nuclei (CDCC),  based on the the random matrix  (RMT)  and the optical background (OPM) models for the statistical theory of nuclear reactions.
The coupled channels equations are divided into two sets. The first set, described by the CDCC, and the other set treated with RMT. The resulting theory is a Statistical CDCC (CDCC$_S$), able in principle to take into account many pseudo channels.
}
\section{Statistical Continuum Discretized Coupled Channels}
\label{intro}
Continuum-Discretized Coupled-Channels (CDCC) calculations are a major theoretical tool to calculate observables in reactions involving rare loosely-bound nuclear isotopes \cite{BC92,DH13}. Such calculations are time-consuming and may include such a huge humber of channels that they are amenable to a statistical treatment, similar to what has been used to treat neutron-induced reactions with compound nuclear states. This is the subject of the present work.

We write the CDCC equations in a schematic model as,
\begin{equation}
\left [-\frac{\hbar^2}{2\mu}\frac{d^2}{dx^2} + V_{rel}(x) + \epsilon_n - E \right]\psi_n(x)  + \sum_m V_{nm}(x) \psi_m(x) = 0
\end{equation}

We distinguish the desired channels in a conventional CDCC by the labels $m, n$, from the statistical channels labeled by $\mu, \nu$. Accordingly,
\begin{equation}
\left[-\frac{\hbar^2}{2\mu}\frac{d^2}{dx^2} + V_{rel}(x) + \epsilon_n - E \right]\psi_n(x)  + \sum_m V_{nm}(x) \psi_m(x) + \sum_{\mu} V_{n\mu}(x) \psi_{\mu}(x)  = 0
\end{equation}
and
\begin{equation}
\left[-\frac{\hbar^2}{2\mu}\frac{d^2}{dx^2} + V_{rel}(x) + \epsilon_{\mu} - E \right]\psi_{\mu}(x)  + \sum_{\nu} V_{\mu\nu}(x) \psi_{\nu}(x) + \sum_{m} V_{\mu m}(x) \psi_{m}(x)  = 0
\end{equation}

\begin{figure}
\centering
    \includegraphics[width=0.5\textwidth]{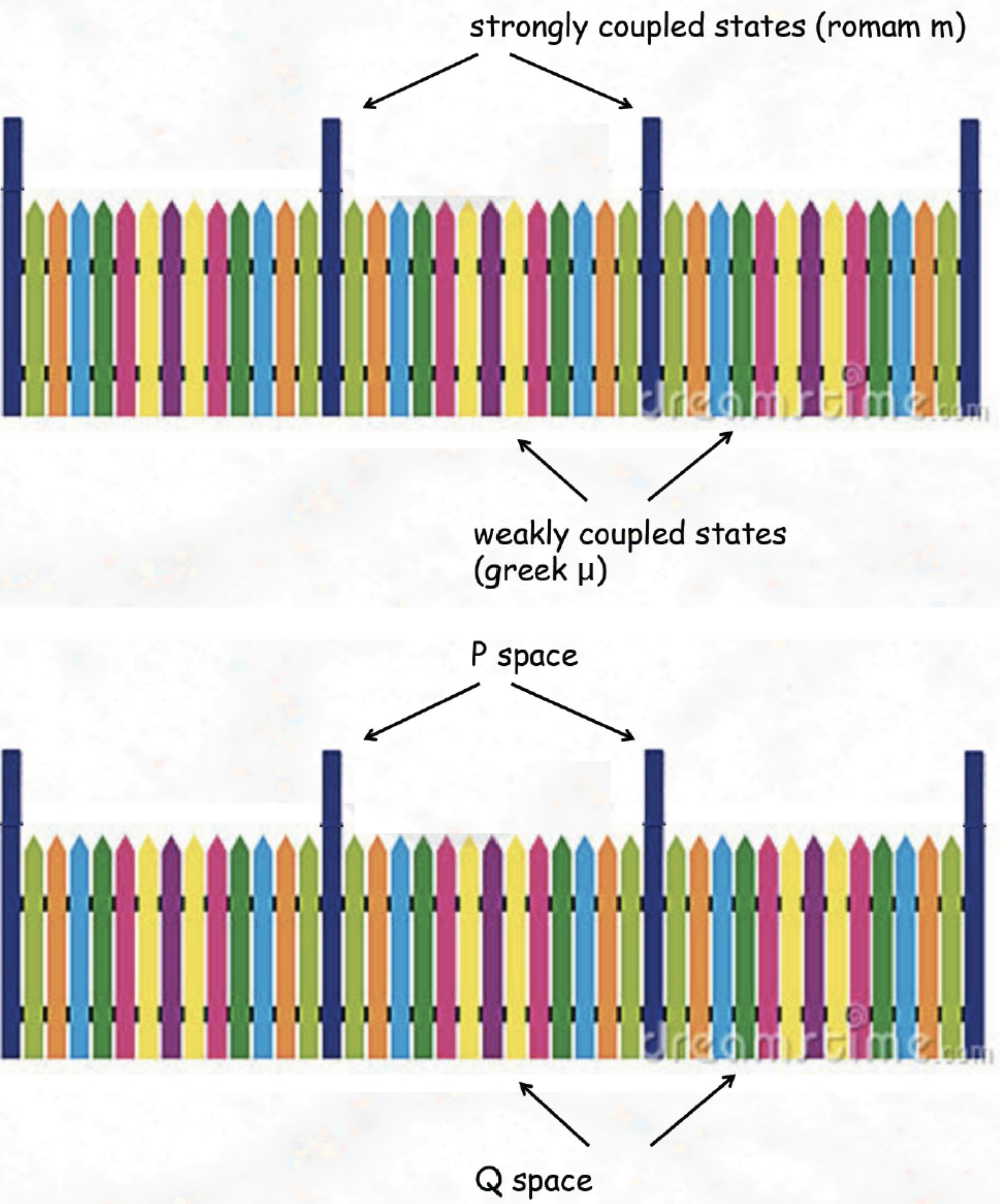}
  \caption{Strongly coupled states belonging to space P and denoted by roman letters, are weakly coupled to states belonging to space Q and denoted by greek letters.}
  \label{fig1}
\end{figure}

The statistical nature of the $\mu, \nu$ channels is specified through the following properties of the matrix elements,
\begin{equation}
\overline{V_{\mu \nu}} = 0
\end{equation}
\begin{equation}
\overline{V_{\mu \nu} V_{\eta \rho}} = (\delta_{\mu, \eta}\delta_{\nu, \rho} + \delta_{\mu, \rho}\delta_{\nu, \eta})\overline{V_{\mu, \nu}^2}
\end{equation}
where the second moment $\overline{V_{\mu, \nu}^2}$ can be parametrized as,
\begin{equation}
\overline{V_{\mu, \nu}^2} = \frac{\omega_0}{\sqrt{\rho(\epsilon_{\mu})\rho(\epsilon_{\nu})}} \exp\left\{-\frac{(\epsilon_{\mu} + \epsilon_{\nu})^2}{2\Delta^2}\right\}
\end{equation}
with $\rho(\epsilon)$ being the density of states, and $\omega_0$ and $\Delta$ are adjustable parameters. 

The same statistical properties are invoked on the matrix elements $V_{\mu, m}$, etc.
\begin{equation}
\overline{V_{\mu m}} = 0,
\end{equation}
\begin{equation}
\overline{V_{\mu m} V_{\eta n}} = (\delta_{\mu, \eta} \delta(m, n)+ \delta_{\mu, n}\delta_{n, \eta})\overline{V_{\mu, m}^2},
\end{equation}
where the second moment $\overline{V_{\mu, m}^2}$ can be parametrized as,
\begin{equation}
\overline{V_{\mu, m}^2} = \frac{\omega_0}{\rho(\epsilon_{\mu})} \exp\left\{-\frac{(\epsilon_{\mu} + \epsilon_{\eta})^2}{2\Delta_{m}^2}\right\}
\end{equation}
with $\rho(\epsilon)$ being the density of states, and $\omega_0$ and $\Delta$ are adjustable parameters. \\

The above prescription should set the stage for a CDCC calculation which presents fluctuations in the final channels and one must rely on an appropriate ensemble average:
perform the calculation several times and at the end perform an average.

\section{CDCC$_S$ in the random matrix formalism}
\label{sec-1}
It is natural to expect that the CDCC equations for the $m, n$ channels would be affected by the statistical channels $(\mu, \nu)$. Averaging the equations above is rather difficult. An easier procedure is to eliminate the statistical channels in favor of the CDCC channels $(m, n)$, resulting in 
\begin{equation}
\psi_{\mu}(x) = \int_{0}^{\infty} dx^{\prime} G_{\mu}(x, x^{\prime})\sum_{m} V_{\mu m}(x^{\prime}) \psi_{m}(x^{\prime})   
\end{equation}
where $G_{\mu}(x, x^{\prime})$ is the diagonal elements of a matrix Green's function defined through the equation
\begin{equation}
\left[[-\frac{\hbar^2}{2\mu}\frac{d^2}{dx^2} + V_{rel}(x) +  \epsilon_{\mu} - E]\delta_{\mu,\nu} +  V_{\mu\nu}(x)  \right]G_{\mu, \nu}(x, x^{\prime}) = \delta( x - x^{\prime})
\end{equation}

With the above Green's function we have for the effective CDCC equations,
\begin{equation}
\left[-\frac{\hbar^2}{2\mu}\frac{d^2}{dx^2} + V_{rel}(x) + V_{pol, n}(x) + \epsilon_n - E \right]\psi_{n}(x)  + \sum_m V_{nm}(x) \psi_{m}(x)  = 0
\end{equation}
where $V_{pol, n}(x)$, is given by
\begin{equation}
V_{pol, n}(x)\psi_{n}(x) = \int_{0}^{\infty}dx \sum_{\mu} V_{n\mu}(x) \int_{0}^{\infty} dr^{\prime} G_{\mu}(x, x^{\prime})\sum_{m} V_{\mu m}(x^{\prime}) \psi_m(x^{\prime})
\end{equation}

The polarization potential fluctuates owing to the random nature of the coupling $V_{n, \mu}$. Thus we have to average this equation over the ensemble. What remains is the fluctuation contribution which we address later.  Thus,
\begin{equation}
\overline{V_{pol, n}(x)\psi_{n}(x)} = \sum_{\mu} \overline{V_{n\mu}(x) \int_{0}^{\infty} dx^{\prime} G_{\mu}(x, x^{\prime})\sum_{m} V_{\mu m}(x^{\prime}) \psi_m(x^{\prime})}.
\end{equation}

The above equation is difficult to average. What we can do is to borrow from the Optical Background Representation of KKM, and introduce the average polarization potential,
\begin{equation}
\overline{V_{pol, n}(x)} = \sum_{\mu} \overline{V_{n\mu}(x) \int_{0}^{\infty} dx^{\prime} G_{\mu}(x, x^{\prime})\sum_{m} V_{\mu m}(x^{\prime})}
\end{equation}

To perform the average above, we have to expand the Green's function in $V_{\mu \nu}$ and then consider only even powers of V,s. This is quite lengthy and was done by Weidenm\"uller. Here we ignore the fluctuations in G, and proceed to average $\overline{V_{n\mu}(x)V_{\mu m}(x^{\prime})} = [\delta_{n,\mu}\delta_{\mu,m} + \delta_{n,m}] F(x,x^{\prime}) \overline{V_{\mu, m}^2}$, where $F(x, x^{\prime}) = \exp\left\{-(x - x^{\prime})^2/2\sigma^2\right\}$.

Accordingly, we have the CDCC$_S$ equations with the average polarization potential read,
\begin{equation}
\left[-\frac{\hbar^2}{2\mu}\frac{d^2}{dx^2} + V_{rel}(x) + \overline{V_{pol, n}(x)} + \epsilon_n - E \right]\overline{\psi_{n}(x)} + \sum_m V_{nm}(x) \overline{\psi_{m}(x)}  = 0.
\end{equation}

The above equations are the new CDCC ones appropriate for our purpose. The fluctuation contribution can be obtained by going back to the original equation and write
$V_{pol, n}(x) = \overline{V_{pol, n}(x)} + V_{pol, n}^{fl}(x)$. This implies that the CDCC wave functions themselves fluctuate. However, the CDCC equations are not the appropriate venue to obtain the fluctuation contribution. What we need is an equation for the square of the CDCC wave functions: a master equation. This was obtained by Weidenm\"ulcer \cite{Ag},
and \cite{Ko}. We should mention that fluctuations in the final state has been discussed in the case of transfer leading to the excitation of an isobaric analog resonance \cite{KM}.
\section{Time-dependent CDCC$_S$}

In the c.m. of the projectile, we assume
$H=H_{0}+V$,
where $H$\ is  composed by the
non-perturbed Hamiltonian $H_{0}$\ and a small perturbation $V$. The
Hamiltonian $H_{0}$\ satisfies an eigenvalue equation
$
H_{0}\psi_{n}=E_{n}\psi_{n},\label{hove}%
$
whose eigenfunctions form a complete basis (including continuum) in which the total wavefunction
$\Psi$, that obeys
$
H\Psi=i\hbar{\partial\Psi}/{\partial t},\label{eqschro}%
$
can be expanded as
$
\Psi=\sum_{n}a_{n}(t)\psi_{n}e^{-iE_{n}t/\hbar}.\label{exppsi}%
$
One obtains
\begin{equation}
i\hbar\sum_{n}\dot{a}_{n}\psi_{n}e^{-iE_{n}t/\hbar}=\sum_{n}Va_{n}\psi
_{n}e^{-iE_{n}t/\hbar},\label{5.42}%
\end{equation}
with $\dot{a}_{n}\equiv d a_{n}(t)/d t$. Using the
orthogonalization properties of the $\psi_{n}$, let us multiply (\ref{5.42})
by $\psi_{k}^{\ast}$\ and integrate it in the coordinate space. From this,
results the the time-dependent coupled channels equations%
\begin{equation}
\dot{a}_{k}\left(  t\right)  =-\frac{i}{\hbar}\sum_{n}a_{n}\left(  t\right)
V_{kn}\left(  t\right)  \ e^{i{\frac{E_{k}-E_{n}}{\hbar}}t},\label{5.43}%
\end{equation}
where we introduced the matrix element ($d\tau$\ is the volume element)
$
V_{kn}=\int\psi_{k}^{\ast}V\psi_{n}\,d\tau.\label{5.44}%
$

\begin{figure}
\centering
    \includegraphics[width=0.5 \textwidth]{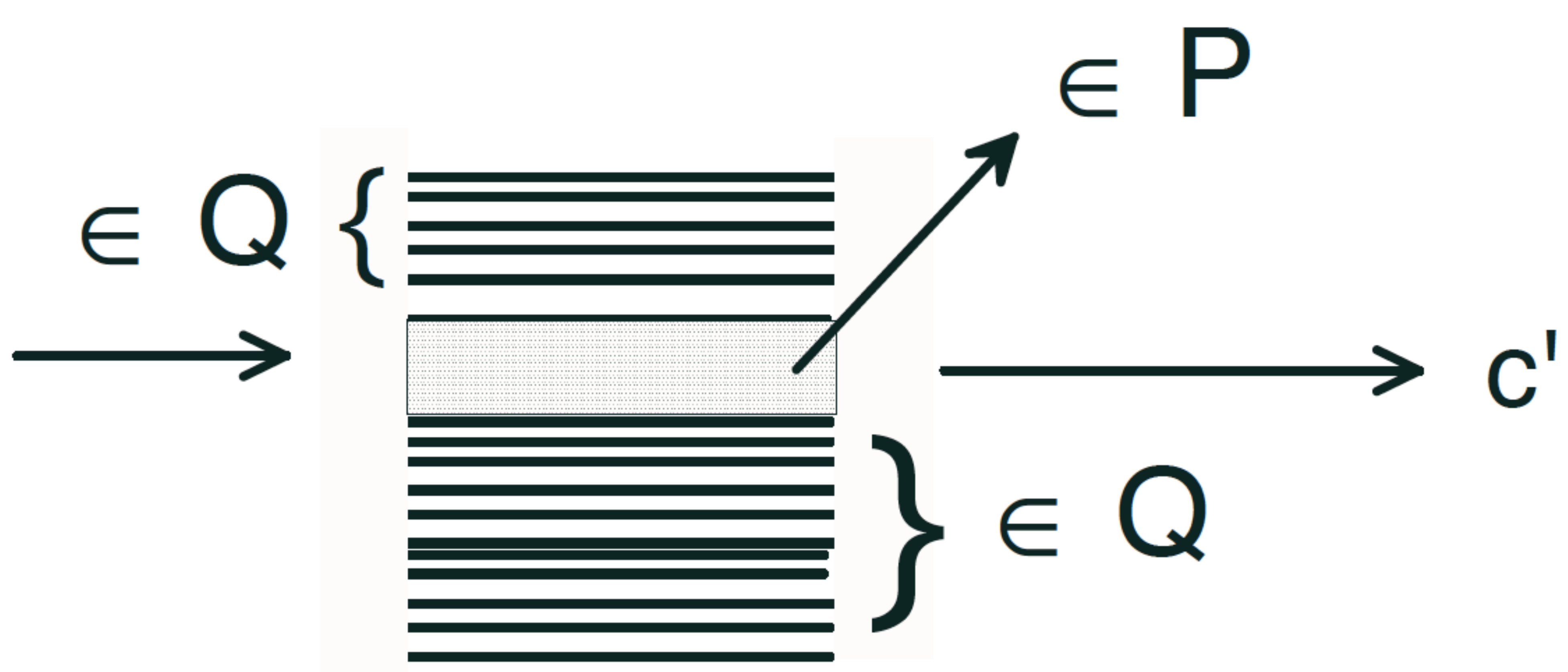}
  \caption{Scematic representation of the P and Q spaces in the optical background formalism.}
  \label{fig2}
\end{figure}

In order to get balance equations, we follow a method described in Ref. \cite{BB88}  (section 3.2.3) for the excitation of giant resonances in relativistic heavy Ion collisions. We rewrite Eq. \eqref{5.43} with explicit account of P (bound + ``strongly-coupled states" in continuum - denoted by roman letters) + Q (``weakly-coupled" states in the continuum - denoted by greek letters)
\begin{equation}
\dot{a}_{k}\left(  t\right)  =-\frac{i}{\hbar}\sum_{n}a_{n}\left(  t\right)
V_{kn}\left(  t\right)  \ e^{i{\frac{E_{k}-E_{n}}{\hbar}}t}-\frac{i}{\hbar}\sum_{n}a_{\mu}\left(  t\right)
V_{k\mu}\left(  t\right)  \ e^{i{\frac{E_{k}-E_{\mu}}{\hbar}}t} +\int_{-\infty}^\infty dt' K_k(t-t')a_k(t'),\label{eqball}%
\end{equation}
where we introduced the last term to account for decay to other channels than the breakup channel under scrutiny. This function is given in terms of the width of the state $k$ as
\begin{equation}
K(t-t')=-{i\over 4\pi}\int_{-\infty}^\infty d\omega\ e^{i\omega (t-t')}\ \Gamma_k\left( \omega-{E_k\over \hbar}\right).\label{eqball2}%
\end{equation}
For $\Gamma_k$ equal to a constant, one obtains $K_k(t-t')=-i(\Gamma_k/2)\delta(t-t')$, we get
\begin{equation}
\dot{a}_{k}\left(  t\right)  =-\frac{i}{\hbar}\sum_{n}a_{n}\left(  t\right)
V_{kn}\left(  t\right)  \ e^{i{\frac{E_{k}-E_{n}}{\hbar}}t}-\frac{i}{\hbar}\sum_{n}a_{\mu}\left(  t\right)
V_{k\mu}\left(  t\right)  \ e^{i{\frac{E_{k}-E_{\mu}}{\hbar}}t} -i{\Gamma_k \over 2} a_k(t).\label{eqball2}%
\end{equation}
Since $\Gamma_k  \neq 0$, the total probability $P = \sum_k |a_k(t)|^2$ is no longer conserved because a flux is now put into the decay channels. Multiplying Eq. \eqref{eqball2} by $a_k^\ast $ and its complex conjugate by $a_k(t)$ and subtracting the results, we obtain for the occupation probability $P_k(t)=|a_k(t)|^2$,
\begin{equation}
\dot P_{k}\left(  t\right)  =\frac{2}{\hbar}\Im  \left[ \sum_{n}a_{n}^\ast (t) a_k(t)
V_{kn}\left(  t\right)  \ e^{i{\frac{E_{k}-E_{n}}{\hbar}}t}\right] +\frac{2}{\hbar}\Im  \left[ \sum_{n}a_k^\ast (t) a_{\mu}\left(  t\right)
V_{k\mu}\left(  t\right)  \ e^{i{\frac{E_{k}-E_{\mu}}{\hbar}}t}\right] -{\Gamma_k \over \hbar} P_k(t).\label{eqball3}%
\end{equation}
This equation can be written as a balance equation for the probability in the form
\begin{equation}
\dot P_{k}\left(  t\right)  =G_k(t)+L_k(t),\label{eqball4}%
\end{equation}
where the gain term is obtained whenever
\begin{equation}
H_k(t)= \frac{2}{\hbar}\Im \left[ \sum_{n}a_{n}^\ast (t) a_k(t)
V_{kn}\left(  t\right)  \ e^{i{\frac{E_{k}-E_{n}}{\hbar}}t}\right] +\frac{2}{\hbar}\Im  \left[ \sum_{n}a_k^\ast (t) a_{\mu}\left(  t\right)
V_{k\mu}\left(  t\right)  \ e^{i{\frac{E_{k}-E_{\mu}}{\hbar}}t}\right] ,\label{eqball5}
\end{equation}
is positive, i.e., $G_k^{(+)}= {\rm positive}(H_k)$, and the loss term is obtained whenever $H_k(t)$ is negative added to the loss by decay,
\begin{equation}
L_{k}\left(  t\right)  =G^{(-)}_k(t) -{\Gamma_k \over \hbar} P_k(t).\label{eqball6}%
\end{equation}

Similar equations are obtained for the occupation probability $P_\mu$ due to $n\mu$ (PQ) and $\nu\mu$ (QQ) coupling. The first part on the rhs of Eq. \eqref{eqball5} is obtained directly from solving the cc equations with the ``strongly interacting" P states. But the second term has to be averaged in a proper way. The same needs to be done for both terms of $H_\mu$ which also contains terms proportional to $a_\mu a_\nu$. If the averaging is possible then the cc equation can be reiterated to include the couplings involving the continuum till convergence is achieved.

\section{Optical background CDCC$_S$}
We consider the scattering of a projectile with center of mass energy $E$, which by means of the interaction with the target and between the core and neutron (say, for $^{11}$Be $\rightarrow$ $^{10}$Be+n) may  transit to other channels.   We consider scattering within a narrow band of discretized continuum states  belonging to a space P. All other states, continuum + bound states, will be assumed to belong to space Q. We emphasize here that this decomposition of the Hilbert space is done on the final wave function which contains the breakup channels. As such the energy mentioned above should be taken as that pertaining to the nucleus which breaks up. If the energy is taken as the total CM one and the decomposition is done on the total wave function, then we are dealing with the conventional compound nucleus reaction, which is not the aim of this investigation. This is an extension to breakup reactions of a formalism developed in Ref. \cite{Arb12}. In the following we take $E$ to be the energy of the subsystem.

The Schr\"odinger equation (SE)  for this problem is
$ H\Psi=E\Psi $, containing 
an  internal (core-neutron) interaction $V$ and  the projectile interaction with the target $U$.   
Using the Feshbach formalism, we introduce the projection operators $P$ and $Q$, so that $Q=1-P$ and
$\Psi=P\Psi+Q\Psi$. 
Then,  for the part of the
wavefunction in space $P$, we get
\begin{equation}
\left(  E-H_{PP}-H_{PQ}G_{Q}H_{QP}\right)  P\Psi=0, \label{PQ2}%
\end{equation}
where
\begin{equation}
G_{Q}\equiv G_{Q}\left(  E\right)  =\frac{1}{E-H_{QQ}+i\varepsilon}.
\label{GQ}%
\end{equation}

\begin{figure}[t]
\centering
    \includegraphics[width=0.65\textwidth]{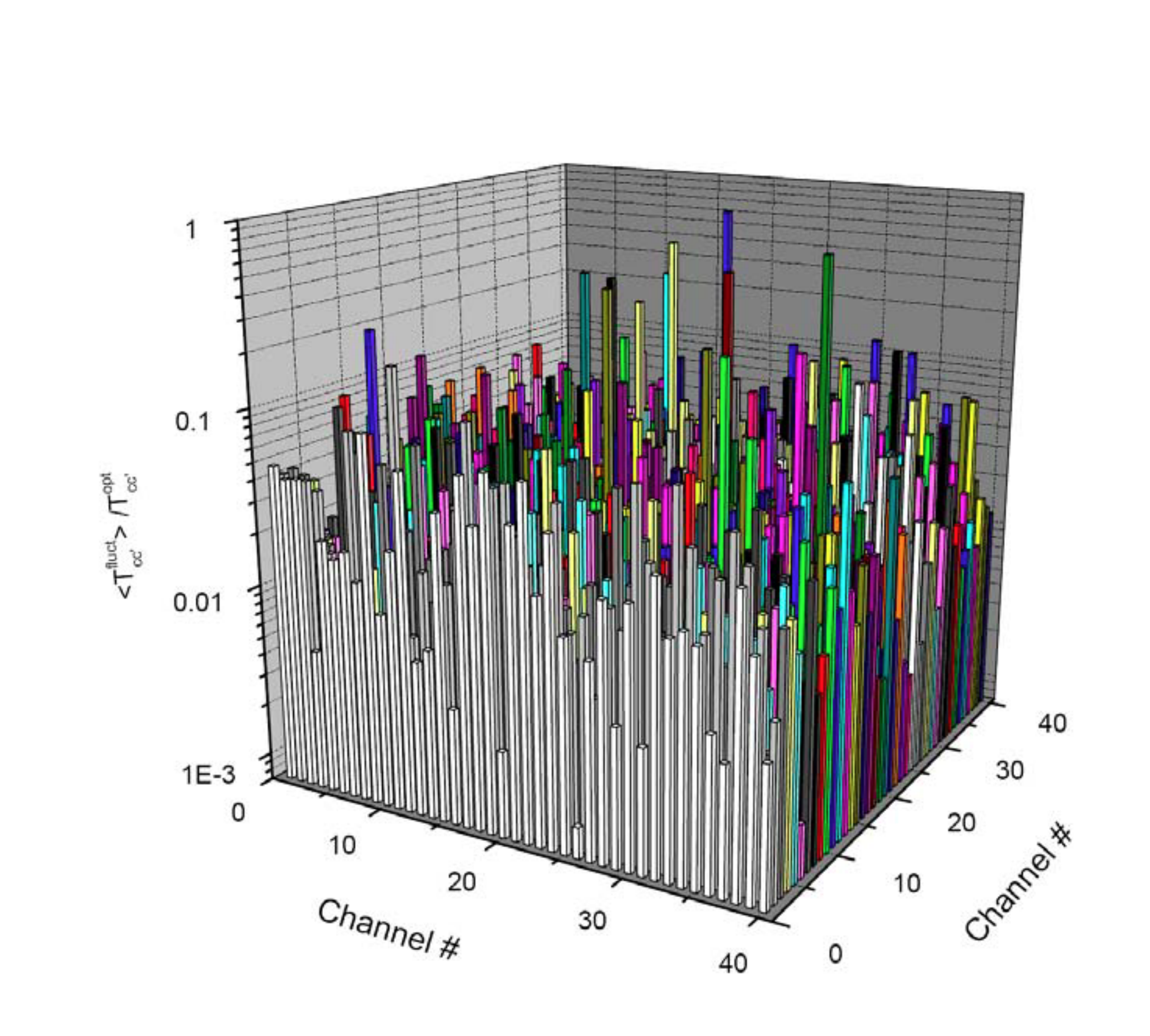}
  \caption{Ratio between the average of the fluctuating part and the optical part of the T-matrix.}
  \label{fig3}
\end{figure}

The continuum is discretized by averaging the 
wavefunction in the channel $P$ is over an energy interval $\Delta$
according to the prescription%
\begin{equation}
\left\langle \Psi\right\rangle _{\Delta}    =\frac{\Delta}{2\pi}\int_{-\infty}^{\infty
}dE^{\prime}\frac{\Psi_{E^{\prime}}}{\left(  E-E^{\prime}\right)  ^{2}%
+\Delta^{2}/4} =\Psi\left(  E+i\frac{\Delta}{2}\right)  , \label{OPT1}%
\end{equation}
where  the residue theorem was used.

It is then straightforward to show that
\begin{equation}
P\Psi={\Psi}_{P}^{\mathrm{\Delta}}+\mathcal{G}_{P}\mathcal{V}_{PQ}\frac
{1}{E-H_{QQ}-\mathcal{V}_{QP}\mathcal{G}_{P}\mathcal{V}_{PQ}}%
\mathcal{V}_{QP}{\Psi}_{P}^{\Delta}, \label{OPT91}%
\end{equation}
with $\Psi_{P}^{\Delta}=\left\langle P\Psi\right\rangle _{\Delta}\simeq P\Psi(E+i\Delta/2)$ being the solution of the energy average of Eq. \eqref{PQ2}, which is obtained by replacing $G_Q(E)$ by $G_Q(E+i\Delta/2)$. This follows from the assumption that solutions of $H_{PP}$ and $H_{PQ}$ are slowly varying
functions of $E$. In Eq. \eqref{OPT91},
\begin{align}
\mathcal{G}_{P}\left(  E\right)   
  =\frac{1}{E-H_{PP}-H_{PQ}G_{Q}\left(  E+i\Delta/2\right)  H_{QP}},
\label{OPT10}%
\end{align}
and 
\begin{align}
\mathcal{V}_{PQ}\left(  E\right)   &  =H_{PQ}\left(  E\right)  \left(
\frac{i\Delta/2}{E+i\Delta/2-H_{QQ}+i\varepsilon}\right)  ^{1/2}\nonumber\\
\mathcal{V}_{QP}\left(  E\right)   &  =\left(  \frac{i\Delta/2}{E+i\Delta/2-H_{QQ}%
+i\varepsilon}\right)  ^{1/2}H_{QP}\left(  E\right)  . \label{OPT8}%
\end{align}

We now consider the Hamiltonian%
\begin{equation}
W=H_{QQ}+\mathcal{V}_{QP}\mathcal{G}_{P}^{opt}\mathcal{V}_{PQ} \label{S1}%
\end{equation}
which describes the coupling of the $Q$ and $P$ spaces, e.g. bound-state-continuum and continuum-continuum couplings.
We assume that a complete set of projectile eigenstates $\left\vert \omega \right\rangle $, with eigenenergies $\mathcal{E}_\omega$
can be found for $W$. 
The eigenvalue problems are
very different: ${\Psi}^{\mathrm{opt}}_{P}$ is solution of $H_{PP}-H_{PQ}%
G_{Q}\left(  E+i\frac{\Delta}{2}\right)  H_{QP}$, whereas $\left\vert
\omega\right\rangle $ is solution of $H_{QQ}+\mathcal{V}_{QP}\mathcal{G}_{P}%
^{opt}\mathcal{V}_{PQ}$. 

The S-matrix, $S_{cc^{\prime}}=\left\langle \left(  P\Psi\right)
_{c}^{\left(  -\right)  }|\left(  P\Psi\right)  _{c^{\prime}}^{\left(
+\right)  }\right\rangle $ can be obtained  using steps similar to the derivation of the
Gell-Mann-Goldberger relation (or two-potential formula). 
One gets 
\begin{equation}
S_{cc^{\prime}}\left(  E\right)  =\overline{S}_{cc^{\prime}}\left(  E\right)
-i\sum_{\omega}\frac{\gamma_{\omega c}\gamma_{\omega c^{\prime}}}{E-\mathcal{E}_{\omega}}, \label{S41}%
\end{equation}
where%
\begin{equation}
\gamma_{\omega c}(E)=\sqrt{2\pi}\left\langle \omega\left\vert \mathcal{V}_{QP}\left(
E\right)  \right\vert {\Psi}^{\Delta}_{P=c}\right\rangle . \label{S5}%
\end{equation}
The channels $c$ and $c^{\prime}$ in eq. \eqref{S41} denote the
different channels for the scattering matrix $S_{cc^{\prime}}$. 

The result above splits the S-matrix into a direct part, $\overline{S}_{cc^{\prime}}\left(  E\right)=\left\langle   \Psi^{\Delta(-)}
_{c}  |\left(  \Psi\right)  _{c^{\prime}}^{\Delta
(+)  }\right\rangle $,  and a multi-step part, the second term in Eq. \eqref{S41}, which contains all the multiple couplings between the $Q$ and $P$ spaces.  Assuming that  $\gamma_{\omega c}$ are smooth functions of $E$, we can perform the ensemble average of the coupling
term in Eq. \eqref{S41} as%
\begin{align}
 \left\langle \sum_{\omega}\frac{\gamma_{\omega c}\gamma_{\omega c^{\prime}}}{E-\mathcal{E}_{\omega}%
}\right\rangle _{\Delta} =
 \sum_{\omega} \frac{\gamma_{\omega c}(E+i\Delta/2)\gamma_{\omega c^{\prime}}(E+i\Delta/2)} {E+i\Delta/2-\mathcal{E}_\omega}. \label{aveg}%
\end{align}

In the channel $P,$ we define an continuum-discretized model wave function%
\begin{equation}
\overline{\Psi}_{P}=\Psi_{P}^{opt}=\left\langle P\Psi\right\rangle _{\Delta}\simeq
P\Psi_{E+i\Delta/2}, \label{OPT2}%
\end{equation}
where
${\Psi}^{\mathrm{opt}}_{P}$ is the solution of
\begin{equation}
\left[  E-H_{PP}-H_{PQ}G_{Q}\left(  E+i\frac{\Delta}{2}\right)  H_{QP}\right]
{\Psi}^{\mathrm{opt}}_{P}=0, \label{OPT3}%
\end{equation}
or $\left(  E-\mathcal{H}_{P}^{opt}\right)  {\Psi}^{\mathrm{opt}}_{P}=0,$ with%
\begin{equation}
\mathcal{H}_{P}^{opt}\left(  E\right)  =H_{PP}+H_{PQ}G_{Q}\left(  E+i\frac
{\Delta}{2}\right)  H_{QP}. \label{OPT4}%
\end{equation}
Eq. \eqref{OPT3} follows from an average of eq. \eqref{PQ2}, with the assumption
that solutions of $H_{PP}$ and $H_{PQ}$ are slowly varying functions of $E$.

Using the simple relation%
\begin{equation}
G_{Q}(E)-G_{Q}\left(E+i\frac{\Delta}{2}\right)=G_{Q}(E)\frac{i\Delta}{2}G_{Q}\left(E+i\frac{\Delta}{2}\right),
\label{OPT5}%
\end{equation}
and from eq. \eqref{PQ2}, by adding and subtracting $\mathcal{H}_{P}^{opt}$, one gets
\begin{equation}
\left[  E-\mathcal{H}_{P}^{opt}-\mathcal{V}_{PQ}\left(  E\right)
G_{Q}(E)\mathcal{V}_{QP}\left(  E\right)  \right]  P\Psi=0, \label{OPT7}%
\end{equation}
where $\mathcal{V}_{PQ}$ and $\mathcal{V}_{QP}$ are given by Eqs. \eqref{OPT8}.

Now we solve eq. \eqref{OPT7} and get 
\begin{equation}
P\Psi={\Psi}_{P}^{\mathrm{opt}}+\mathcal{G}_{P}^{opt}\mathcal{V}_{PQ}\frac
{1}{E-H_{QQ}-\mathcal{V}_{QP}\mathcal{G}_{P}^{opt}\mathcal{V}_{PQ}}%
\mathcal{V}_{QP}{\Psi}_{P}^{\mathrm{opt}}, \label{OPT9}%
\end{equation}
with%
\begin{align}
\mathcal{G}_{P}^{opt}\left(  E\right)     =\frac{1}{E-\mathcal{H}_{P}%
^{opt}\left(  E\right)  }
  =\frac{1}{E-H_{PP}-H_{PQ}G_{Q}\left(  E+i\frac{\Delta}{2}\right)  H_{QP}}.
\label{OPT10}%
\end{align}

The Hamiltonian%
\begin{equation}
W=H_{QQ}+\mathcal{V}_{QP}\mathcal{G}_{P}^{opt}\mathcal{V}_{PQ} \label{S1}%
\end{equation}
describes the coupling of the $Q$ and $P$ spaces, e.g. bound-state-continuum and continuum-continuum couplings and is the basis of the optical background formulation of the CDCC$_S$.

Let us assume that a complete set of projectile eigenstates $\left\vert q\right\rangle $
can be found for $W$. That is%
\begin{equation}
\left[  H_{QQ}+\mathcal{V}_{QP}\mathcal{G}_{P}^{opt}\mathcal{V}_{PQ}\right]
\left\vert q\right\rangle =\mathcal{E}_{q}\left\vert q\right\rangle ,
\label{S2}%
\end{equation}
then we can insert this set into eq. \eqref{OPT9}\ and it becomes%
\begin{equation}
P\Psi={\Psi}^{\mathrm{opt}}_{P}+\mathcal{G}_{P}^{opt}\mathcal{V}%
_{PQ}\left\vert q\right\rangle \frac{1}{E-\mathcal{E}_{q}}\left\langle
q\right\vert \mathcal{V}_{QP}\left\vert {\Psi}^{\mathrm{opt}}_{P}\right\rangle
. \label{S3}%
\end{equation}

The S-matrix, $S_{cc^{\prime}}=\left\langle \left(  P\Psi\right)
_{c}^{\left(  -\right)  }|\left(  P\Psi\right)  _{c^{\prime}}^{\left(
+\right)  }\right\rangle $ can be obtained from the equation above and its
complex conjugate, using steps similar to the derivation of the
Gell-Mann-Goldberger relation (or two-potential formula). The Hamiltonians are
very different: ${\Psi}^{\mathrm{opt}}_{P}$ is solution of $H_{PP}-H_{PQ}%
G_{Q}\left(  E+i\frac{\Delta}{2}\right)  H_{QP}$, whereas $\left\vert
q\right\rangle $ is solution of $H_{QQ}+\mathcal{V}_{QP}\mathcal{G}_{P}%
^{opt}\mathcal{V}_{PQ}$. One gets 
\begin{equation}
S_{cc^{\prime}}\left(  E\right)  =\overline{S}_{cc^{\prime}}\left(  E\right)
-i\sum_{q}\frac{g_{qc}g_{qc^{\prime}}}{E-\mathcal{E}_{q}}, \label{S4}%
\end{equation}
where%
\begin{equation}
g_{qc}(E)=\sqrt{2\pi}\left\langle q\left\vert \mathcal{V}_{QP}\left(
E\right)  \right\vert {\Psi}^{\mathrm{opt}}_{P=c}\right\rangle . \label{S5}%
\end{equation}

We first consider the eigenvalues of the operator
\begin{equation}
W_{0}=H_{QQ}-H_{QP}G^{0}H_{PQ},\label{W}%
\end{equation}
where $G^{0}$ is a free projectile Green's function
\begin{equation}
G^{0}=\frac{1}{E-H_{PP}},\label{gfree}%
\end{equation}
and where a single particle projection operator $P$ in spatial representation
is
\begin{equation}
P=\sum_{c}\int r^{2}dr|r,c\rangle\langle r,c|\equiv1-Q\label{P}%
\end{equation}
where, for the time being, $|r,c\rangle$ is to be distinguished from the
initial (or final) free projectile wave-function $|\phi_{i}\rangle$ (or
$|\phi_{f}\rangle$)
\begin{equation}
(E_{i}-H_{PP})|\phi_{i}\rangle=0,
\end{equation}
with%
\begin{equation}
\left\langle r,c|r^{\prime},c^{\prime}\right\rangle =\frac{\delta\left(
r-r^{\prime}\right)  }{rr^{\prime}}\delta_{cc^{\prime}}.\label{norm1}%
\end{equation}
Eq.
\eqref{W} can be written with intrinsic nuclear state indices displayed explicitly as
\begin{equation}
W_{jk}\equiv\langle Q_{j}|W_{0}|Q_{k}\rangle=\delta_{jk}E_{j}^{(Q)}%
+H_{jP}G^{0}H_{Pk},\label{wjk}%
\end{equation}
where the second term above can be expanded by virtue of operator $P$ in Eq.
\eqref{P} as
\begin{equation}
H_{jP}G^{0}H_{Pk}=\sum_{cc^{\prime}}\int r^{2}dr\int{r^{\prime}}^{2}%
dr^{\prime}H_{jc}(r)G_{cc^{\prime}}^{0}(r,r^{\prime})H_{c^{\prime}k}%
(r^{\prime})\label{HGH}%
\end{equation}
where $H_{cj}(r)$ is
\begin{equation}
H_{cj}(r)=\langle r,c|H|Q_{j}\rangle.
\end{equation}
It is now convenient to define the matrix%
\begin{equation}
M_{jk}\left(  E_{c}\right)  =H_{jP}G^{0}\left(  E_{c}\right)  H_{Pk}%
,\label{mjk}%
\end{equation}
in terms of which, eq. \eqref{wjk} can be cast into the form%
\begin{equation}
W_{jk}=\delta_{jk}E_{j}^{(Q)}+M_{jk}.\label{wjk1}%
\end{equation}

Matrix $M_{jk}$ can be conveniently separated into its principal value and an imaginary part
\begin{eqnarray}
M_{jk}\left(  E \right)  = H_{jP} \frac{{\cal P}}{E-H_{PP}}  H_{Pk}  - i \pi  H_{jP} \delta (E-H_{PP})  H_{Pk} 
                      \equiv& D_{jk}(E)  - i {\Gamma_{jk}(E)}/{2},\label{mjk3}
\end{eqnarray}
where matrices $D$ and $\Gamma$ are defined for convenience, in a notation that alludes to resonance shifts and widths that will be obtained
by diagonalizing $W_{jk}$.
A completeness relation, $\sum_c \int dE |\chi_{E;c}^{^{(+)}}\rangle \langle\chi_{E;c}^{^{(+)}}| = 1$, 
for eigenstates of $H_{PP}$ can be used to write $\Gamma_{jk}$ as
\begin{eqnarray}
\Gamma_{jk}\left(  E \right)  =  2\pi\sum_c H_{jP} |\chi_{E;c}^{^{(+)}}\rangle \langle\chi_{E;c}^{^{(+)}}|  H_{Pk}, \label{Gjk1} 
                              =  2\pi\sum_{c} \int_{rr'} H_{jc}(r) \chi^{^{(+)}}_{E;c}(r) \chi_{E;c}^{^{(+)}*}(r')  H_{ck}(r'),\nonumber \label{Gjk2}
\end{eqnarray}
where for a free particle 
$\langle r;c|\chi_{E;c'}^{^{(+)}}\rangle \equiv \chi^{^{(+)}}_{E;cc'}(r)  = \delta_{cc'}\chi^{^{(+)}}_{E;c}(r)$.

A dispersion relation between the real and imaginary part of $M_{jk}$ in Eq. \eqref{mjk3} yields
\begin{equation}
D_{jk}(E)  = \frac{1}{2\pi} {\cal P} \int_0^\infty \frac{\Gamma_{jk}(E')dE'}{E-E'}.\label{djk}
\end{equation}
The energy dependence of $D_{jk}(E)$ will come from $\Gamma_{jk}\left(  E \right)$ and from asymmetric limits of integration.  
Computation of Green's function matrix is thus simplified as it depends on (approximately) free-particle eigenfunctions alone.

\begin{figure}[h]
\centering
    \includegraphics[width=0.6\textwidth]{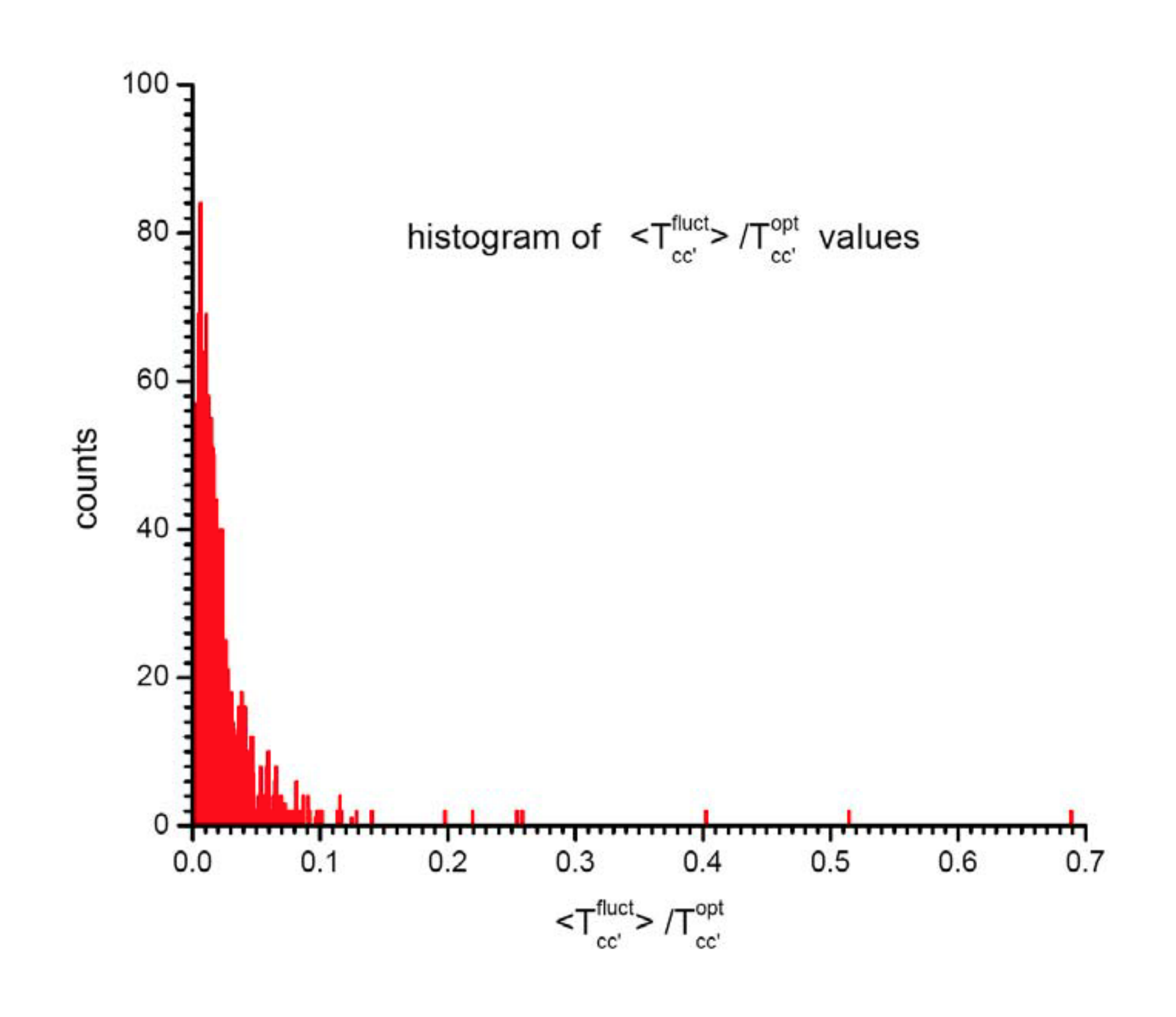}
  \caption{Number of ``counts" for the ratio between the average of the fluctuating part and the optical part of the T-matrix.}
  \label{fig4}
\end{figure}

We write
\begin{equation}
H_{cj}(r)=\sum_{k}h_{cjk}\frac{\delta(r-r_{k})}{rr_{k}},\label{int}%
\end{equation}
where $h_{cjk}$ are complex numbers obtained from the  interaction at projectile position $r_k$, sandwiched between states $j$ and $c$. That is,  the label $k$ in the couplings $h_{cjk}$ denote the location of the
spatial coordinate where the interaction \eqref{int} assumes a non-zero value.
The label $c$,
denotes the channel index. As we are not including any other quantum number
than the energy of the channel, the discretization of energy is the same as
the label $c$. That is, $c$ labels energies varying within the an interval of
width $\Delta$ (the same energy interval used in integral \eqref{OPT1}). The label
$j$ denotes the Q-space. In our case, it will be attached to the energy,
$E_{Q_{j}}\equiv E_{j}$.  The matrix elements between channels
$c$ and $c^{\prime}$ will not be needed.
These are included in $H_{PP}$ and are part of the T-matrix we are after.

We now define the T-matrix $T_{cc^{\prime}}$ as
\begin{equation}
T_{if}=T_{if}^{0}+\langle\phi_{i}|H_{PQ}\frac{1}{E-H_{QQ}-H_{QP}G^{0}H_{PQ}%
}H_{QP}|\phi_{f}\rangle, \label{Tif}%
\end{equation}
and we diagonalize the denominator of Eq. \eqref{Tif} in terms of the
eigenfunctions $|\widehat{q}\rangle$ of $W_{0}$,
\begin{equation}
(\widehat{\mathcal{E}}_{q}-W_{0})|\widehat{q}\rangle=0,
\end{equation}
and use that the $|\widehat{q}\rangle$'s are linear combinations of
eigenstates of $H_{QQ}$, $(E_{j}^{(Q)}-H_{QQ})|Q_{j}\rangle=0$
\begin{equation}
|\widehat{q}\rangle=\sum_{j}\widehat{C}_{jq}|Q_{j}\rangle.
\end{equation}
Then we can rewrite eq. \eqref{Tif} as\footnote{We are going to use only the
kinetic energy term for the scattering waves. In this case, $T_{if}^{0}=0$.}
\begin{equation}
T_{if}=T_{if}^{0}+\sum_{qjk}\widehat{C}_{jq}\widehat{H}_{ij}\frac
{1}{E-\widehat{\mathcal{E}}_{q}}\widehat{C}_{kq\ }\widehat{H}_{kf}, \label{T1}%
\end{equation}
where we defined
\begin{align}
\widehat{H}_{ij}   \equiv\langle\phi_{i}|PH|Q_{j}\rangle
  =\sum_{c}\int r^{2}dr\phi_{ic}^{\ast}(r)\langle r,c|H|Q_{j}\rangle
  =\sum_{c}\int r^{2}dr\phi_{ic}^{\ast}(r)H_{cj}(r). \label{OffDiag3}
\end{align}

We now derive
an analogous expression for the $T$-matrix, with $\overline{\Psi_{P}}$ defined
in eq. \eqref{OPT2}. (To simplify notations we use here $\mathcal{G}%
\equiv\mathcal{G}^{opt}$):
\begin{equation}
T_{if}=\overline{T}_{if}+\langle\overline{\Psi}_{i}|\mathcal{V}_{PQ}\frac
{1}{E-H_{QQ}-\mathcal{V}_{QP}\mathcal{G}\mathcal{V}_{PQ}}\mathcal{V}%
_{QP}|\overline{\Psi}_{f}\rangle,\label{Tcc2}%
\end{equation}
by using the following expression for the optical Green's function
$\mathcal{G}$ and the continuum discretized ket $|\overline{\Psi}_{c}\rangle
$:
\begin{eqnarray}
\mathcal{G}   &=&G^{0}
  +G^{0}H_{PQ}\frac{1}{E-H_{QQ}-H_{QP}G^{0}H_{PQ}+i\Delta/2}H_{QP}G^{0}
|\overline{\Psi}_{f}\rangle \nonumber \\&  =&|\phi_{f}\rangle\label{Psiopt}
  +G^{0}H_{PQ}\frac{1}{E-H_{QQ}-H_{QP}G^{0}H_{PQ}+i\Delta/2}H_{QP}|\phi_{f}%
\rangle .
\end{eqnarray}
Instead of diagonalizing $W_{0}$ as above, we will
diagonalize $W$, eq. \eqref{S1},
\begin{equation}
W=H_{QQ}+\mathcal{V}_{QP}\mathcal{G}\mathcal{V}_{PQ}%
\end{equation}
or in matrix notation
\begin{equation}
\mathcal{W}_{jk}=\delta_{jk}E_{j}^{(Q)}+\mathcal{V}_{jP}\mathcal{G}%
\mathcal{V}_{Pk}\label{wjk2}%
\end{equation}
where the second term can be expanded in terms of eigenvalues $|\widehat
{q}\rangle$ of $W_{0}$ we already found 
\begin{eqnarray}
{\cal V}_{jP} {\cal G} {\cal V}_{Pk} &=& {\cal V}_{jP} G^0 {\cal V}_{Pk} +
{\cal V}_{jP} G^0 H_{PQ} \frac{1}{E-H_{QQ}- H_{QP}G^0 H_{PQ} + i\Delta/2} H_{QP} G^0 {\cal V}_{Pk}\\
&=& {\cal V}_{jP} G^0 {\cal V}_{Pk} + \sum_{qj'k'} \widehat{C}_{j'q}
{\cal V}_{jP} G^0 H_{Pj'} \frac{1}{E- \widehat{{\cal E}}_q + i\Delta/2}
H_{k'P} G^0 {\cal V}_{Pk} \widehat{C}_{k'q}
\end{eqnarray}
where
\begin{equation}
\mathcal{V}_{jP}=\sqrt{\frac{i\Delta}{E-E_{j}^{(Q)}+i\Delta/2}}\sum_{c}\int
r^{2}dr\langle Q_{j}|H|r,c\rangle\langle r,c|.
\end{equation}
Using the definition of $M_{jk}$, eq. \eqref{mjk}, we can rewrite this equation
as%
\begin{align}
\mathcal{V}_{jP}\mathcal{GV}_{Pk}   =\sqrt{\frac{i\Delta}{E-E_{j}^{(Q)}+i\Delta/2}%
}\sqrt{\frac{i\Delta}{E-E_{k}^{(Q)}+i\Delta/2}}
  \times\left\{  M_{jk}+\sum_{qj^{\prime}k^{\prime}}\widehat{C}_{j^{\prime}%
q}M_{jj^{\prime}}\frac{1}{E-\widehat{\mathcal{E}}_{q}+i\Delta/2}M_{k^{\prime}%
k}\widehat{C}_{k^{\prime}q}\right\}  \label{vgv}%
\end{align}
Diagonalizing the denominator of Eq. \eqref{Tcc2} in terms of the
eigenfunctions $|q\rangle$ of $W$,
\begin{equation}
(\mathcal{E}_{q}-{W})|q\rangle=0,
\end{equation}
Eq. \eqref{Tcc2} can be rewritten as
\begin{equation}
T_{if}=\overline{T_{if}}+\sum_{qjk}C_{jq}\widehat{\mathcal{V}}_{ij}\frac
{1}{E-\mathcal{E}_{q}}C_{kq}\widehat{\mathcal{V}}_{kf},
\end{equation}
where we define
\begin{align}
\widehat{\mathcal{V}}_{ij} &  \equiv\langle\overline{\Psi}_{i}|P\mathcal{V}%
|Q_{j}\rangle
  =\sum_{c}\int r^{2}dr\overline{\Psi}_{ic}(r)\langle r,c|\mathcal{V}%
|Q_{j}\rangle
  =\sqrt{\frac{i\Delta}{E-E_{j}^{(Q)}+i\Delta/2}}\sum_{c}\int r^{2}dr\overline{\Psi
}_{ic}(r)H_{cj}(r)\label{vij}%
\end{align}
and where we used the fact that $|q\rangle$'s are linear combinations of
$|Q_{j}\rangle$
\begin{equation}
|q\rangle=\sum_{j}C_{jq}|Q_{j}\rangle.
\end{equation}
It is then straight-forward to recognize that, in Eq. \eqref{S4},
\begin{equation}
g_{iq}=\sqrt{2\pi}\sum_{j}C_{jq}\widehat{\mathcal{V}}_{ij}.\label{giq}%
\end{equation}

The channels $c$ and $c^{\prime}$ in eq. \eqref{S4} denote the
different channels for the scattering matrix $S_{cc^{\prime}}$. We
can perform the energy average of the coupling
term in eq. \eqref{S4} as%
\begin{align}
  \left\langle \sum_{q}\frac{g_{qc}g_{qc^{\prime}}}{E-\mathcal{E}_{q}%
}\right\rangle _{\Delta}& \simeq
 \frac{I}{2\pi}\int_{E-n\Delta}^{E+n\Delta}dE^{\prime}\frac{\sum_{q}%
g_{qc}(E^{\prime})g_{qc^{\prime}}(E^{\prime})/(E^{\prime}-\mathcal{E}_{q}%
)}{\left(  E-E^{\prime}\right)  ^{2}+\Delta^{2}/4}\nonumber\\
&  =\frac{I\Delta}{2\pi} \sum_{n=-N_E}^{N_{E}}\frac{\sum_{q}g_{qc}%
(E+n\Delta )g_{qc^{\prime}}(E+n\Delta )}{(E+n\Delta
-\mathcal{E}_{q})\left[\left( n\Delta \right)
^{2}+\Delta^{2}/4\right]}, \label{aveg}%
\end{align}
where $N_E$ is the number of energy grid points in the continuum,  and $\Delta =n\Delta/N_E$. The integer $n$ defines the
integration limits. This average is therefore dependent on 3
parameters:  $\Delta,$ $N_{E}$, and $n$.

To define the $Q$-space we need the real energies $E_{Q_{j}}\equiv
E_{j}$. The average spacing between the continuum levels, $E_{j}$, will be
called $D$, so that $D=\left\langle E_{j+1}-E_{j}\right\rangle $. We will also
need an energy range for the space, which we take as $E-N_{Q}\Delta\leq E_{_{j}%
}^{(Q)}\leq E+N_{Q}\Delta$. The numbers of the energy levels will be denoted by
$N_{L}$. Thus, in order to define the $Q$-space, we need to define the
$E_{j}^{(Q)}$'s, $N_{L}$, and $N_{Q}$ (see Ref. \cite{Arb12}).

In eq. \eqref{int} the interaction is in our model space given by
$H_{ij}(r)=\sum_{l}h_{ijl}\delta\left(  r-r_{l}\right)/r_lr  $,
where $i$ and $j$ denote the energies (channels) $E_c\equiv
E_{i}=E-E_c^*$, and $E_{j}$ in $P$ and $Q$ space, respectively.
$E_c^*$ is the excitation energy.
The index $l$ denotes the position on the coordinate space where the
interaction is active (the discrete points $r_{l}$).
With that, eq. \eqref{OffDiag3} becomes%
\begin{align}
  \widehat{H}_{ij}=\sum_{c}\int r^{2}dr\phi_{ic}^{\ast}(r)H_{cj}%
(r)
  =\sum_{c=1}^{N_{c}}\sum_{l=1}^{N_{R}}\
  \phi_{ic}^{\ast}(r_{l})h_{cjl}, \label{hij}%
\end{align}
where $N_{c}$ is the number of open channels, and $N_{R}$ is the number of
the $r_{l}$ points in the coordinate mesh.
 Another parameter is needed to
account for the size of the coordinate region where the interaction
is active. We call this $R$ which has the value of a typical nuclear
radius. There will be $N_{R}$ points at positions $r_{l}$ within
$0\leq r_{l}\leq R$. Thus, to calculate the integrals we need the
additional parameters $R$ and the $N_{R}$ values of $r_{l}$.

Similarly, the matrix element in eq. \eqref{HGH} is written as%
\begin{align}
  M_{jk}\left(  E_{c}\right)  =H_{jP}G^{0}H_{Pk}
  =\sum_{c,c^{\prime}=1}^{N_{c}}\sum_{l=1}^{N_{R}}\sum_{l^{\prime}=1}^{N_{R}%
}h_{cjl}h_{c^{\prime}kl^{\prime}}G_{cc^{\prime}}^{0}(r_{l},r_{l^{\prime}}).
\label{hgh}%
\end{align}

The free particle Green's \ function in the $s$-wave channel is given by%
\begin{align}
&  G_{cc^{\prime}}^{0}(E;r,r^{\prime})\equiv G^{0}(E_{c},r,r^{\prime}%
)\delta_{cc^{\prime}}  =\cdots
,\label{mgm}%
\end{align}
where $r_{<}$ ($r_{>}$) is the smaller (larger) of ($r$,$r^{\prime}$). We can
rewrite eq. \eqref{hgh} as%
\begin{align}
  M_{jk}\left(  E_{i}\right) & =\left[  H_{jP}G^{0}H_{Pk}\right]
(E_{i})= \cdots  .\label{hghij}%
\end{align}
This determines the matrix $W_{jk}$ in eq. \eqref{wjk}.

At this point, it is worthwhile to rewrite eq. \eqref{vij} by using
the definition of matrices $M_{jk}$ and $\widehat{H}_{ij}$, eqs. \eqref{mjk} and
\eqref{OffDiag3}, respectively. In these terms%
\begin{align}
\widehat{\mathcal{V}}_{ij} (E) 
=\sqrt{\frac{i\Delta}{E-E_{j}^{(Q)}+i\Delta/2}}\left\{
\widehat{H}_{ik}+\right. 
  \left.  \sum_{kj^{\prime}q}\widehat{C}_{jq}\widehat{H}_{ij}\frac
{1}{E-\widehat{\mathcal{E}}_{q}+i\Delta/2}M_{kj}\widehat{C}_{kq}\right\}  .
\label{vijm}%
\end{align}

Thus, $\widehat{\mathcal{V}}_{ij}$\ can be easily calculated after we obtain
matrices $M_{jk}$ and $\widehat{H}_{ij}$, eqs. \eqref{mjk} and \eqref{OffDiag},
respectively. The same applies to  $\mathcal{V}_{jP}$.%

A numerical test of the optical background formalism to treat a large number of channels statistically was done in Ref. \cite{AB07} with 400 equidistant q-levels,
40 channels, with 20 equidistant coordinate  points where HPQ is set to a Gaussian-distributed random interaction. The energy of the single-out P state
was taken as $E = 20$ MeV, and we included  100 E$^\prime$  points for Lorentzian averaging between 18 and 22 MeV. The adopted value of $\Delta = 0.5$ MeV  and we considered s-wave scattering only, with
$\Gamma/D \gg 1$. Figures \ref{fig3} and \ref{fig4} show the avert of the fluctuating part of the T-matrix and the optical T-matrix. It is evident that for a large number of weakly coupled channels the 
optical background formalism yields a proper treatment of the S- and T-matrices. The advantage of the formalism is that one has only to consider the strongly coupled P-space states, with a simple average needed for the 
weakly coupled states which can be numerous.

\section{Conclusions}
CDDC is an important tool to describe reactions with weakly bound nuclei, for which numerous states in the continuum (resonant and non-resonant) have an important weight for the  reaction cross section.
A drawback of the formalism is that if the number of states involved (channels) are too large, the calculations converge too slowly or might not be feasible with present computer resources.  A possible treatment of a large number of weakly bound states is the use of statistical methods. In this work we have proposed several possibilities to include the average over continuum couplings: (a) perform an average of the coupled-channels equations where continuum-continum couplings are treated by means of statistical ensembles, (b) solve time dependent balance equations with statistical averages, (c) use the optical background formalism based on the Feshbach projection method, or (d) a combination of all methods discussed here. Several applications in quantum optics, atomic and nuclear physics are possible, specially in the area of open quantum or mesoscopic systems. 

\bigskip
We acknowledge beneficial discussions with G. Arbanas and A. Kerman. This work was partially supported by the US-DOE grants DE-SC0004971 and DE-FG02- 08ER41533.

\end{document}